\documentclass[aps,prl,nofootinbib,twocolumn,superscriptaddress]{revtex4}

\usepackage{graphicx}
\usepackage{epsfig}
\usepackage{multirow}

\newcommand{\beq}{\begin{equation}}
\newcommand{\eeq}{\end{equation}}

\newcommand{\eg}{{\it e.g.}}

\newcommand{\bea}{\begin{eqnarray}}
\newcommand{\eea}{\end{eqnarray}}

\newcommand{\gsim}{\lower.7ex\hbox{$\;\stackrel{\textstyle>}{\sim}\;$}}
\newcommand{\lsim}{\lower.7ex\hbox{$\;\stackrel{\textstyle<}{\sim}\;$}}

\def\mysection#1{{\bf #1.} }

\begin{document}

\title{
%124 GeV fermiophobic Higgs mimics the standard model Higgs boson at the LHC \\ or \\
Has a fermiophobic Higgs boson been detected at the LHC?
 }

\author{Emidio Gabrielli}
\affiliation{NICPB, Ravala 10, 10143 Tallinn, Estonia}
\author{Barbara Mele}
\affiliation{INFN, Sezione di Roma, c/o Dipartimento di Fisica, Universit\`a di Roma ÒLa SapienzaÓ, P. le A. Moro 2, I-00185 Rome, Italy}
\author{Martti Raidal}
%\email[]{martti.raidal@cern.ch}
\affiliation{NICPB, Ravala 10, 10143 Tallinn, Estonia}

\date{\today}

\begin{abstract}
We show that, in the present inclusive searches for the  Higgs boson at the LHC, a fermiophobic Higgs mimics  the
standard-model-like Higgs if its mass is around 125~GeV. For that mass the order-of-magnitude reduction of fermiophobic Higgs
production cross sections  is compensated by a corresponding increase in 
the Higgs branching fraction into $\gamma\gamma$, while the $WW^*,$ $ZZ^*,$ $Z\gamma$ signal yields  are predicted to be somewhat  smaller.
The excess seen in the ATLAS and CMS fermiophobic Higgs boson searches in $\gamma\gamma$ channel, including the
exclusive vector-boson-fusion analysis, suggests that the LHC sees a fermiophobic instead of a standard-model-like  Higgs boson. 
 If the Higgs boson turns out to be fermiophobic,  many of our present ideas of new physics should be revised.

\end{abstract}

%\pacs{}
%\keywords{}

\maketitle

\mysection{Motivation}
Proving that the Higgs mechanism~\cite{Englert:1964et,Higgs:1964ia,Higgs:1964pj,Guralnik:1964eu} is the origin of electroweak symmetry breaking
is the main scientific aim of the Large Hardon Collider (LHC). Recently both the ATLAS and CMS experiments have 
published their combined results~\cite{Acomb,Ccomb}  for searches for the standard model (SM) Higgs boson in data collected in 2011.
The new combinations confirm the inconclusive evidence of a SM-like Higgs boson with mass $m_H \approx 125$~GeV.
In both experiments the evidence originates predominantly from the excess in the $H\to \gamma\gamma$ channel~\cite{Agg,Cgg}.  
The CMS paper~\cite{Cgg}  also presents results for the exclusive vector boson fusion (VBF) analyses with forward-dijet tagging
that unexpectedly gives an important contribution to the excess.
ATLAS sees a not very significant excess in $H\to ZZ^*\to 4l$ for the same invariant mass~\cite{AZZ}, while the CMS data in this channel~\cite{CZZ}
as well as in $H\to WW^*$~\cite{CWW} is, within errors, consistent with background.  
The fermionic channels $b\bar b$ and $\tau\bar\tau$ analyzed by the CMS~\cite{Ccomb} are not yet sensitive to the 
SM-like Higgs boson with the collected luminosity.

These results were updated in the Moriond 2012 conference where new results by ATLAS and CMS on searches for a fermiophobic (FP)
Higgs boson in the $\gamma\gamma$ channel were presented. 
Both experiments observe consistently an excess of a $125$~GeV  FP Higgs boson with about $3\sigma$  local significance~\cite{AFP,CMSFP}. 
If these hints will be confirmed, this would imply dramatic consequences on our understanding 
of electroweak symmetry breaking. 

The possibility that (some) Higgses couple at tree level only to the gauge sector and not to fermions, the FP Higgs boson, 
is a well known logical option that can arise as a particular limiting case in models with an extended Higgs sector~\cite{Haber:1978jt,Georgi:1985nv}.
However, more than by a particular theoretical model, 
the interest in the FP Higgs scenario
was triggered 
by the new non-trivial Higgs phenomenology at LEP, Tevatron, SSC and LHC~\cite{Basdevant:1992nb,Barger:1992ty,Bamert:1993ah,Pois:1993ay,Stange:1993ya,Diaz:1994pk,Akeroyd:1995hg,Akeroyd:1998ui}.
A consistent model for one FP Higgs boson as an effective low energy field theory of electroweak symmetry 
breaking was formulated only  recently~\cite{Gabrielli:2010cw,Gabrielli:2011yn}. In this framework the fermion masses, including the top quark  mass,  
must come from a different mechanism, \eg,  from technicolor.

In the context of quantum field theory, a pure FP Higgs boson with vanishing Yukawa couplings is not realistic. Whatever
new physics mechanism is responsible for the fermion masses, at loop level non-vanishing 
Yukawa couplings are induced due to renormalization~\cite{Gabrielli:2010cw}. Although the size of induced Yukawa couplings 
depends on the new physics scale $\Lambda,$ that can be considered as a theoretical uncertainty of the scenario, 
the generic prediction of the  FP Higgs scenario is that the Higgs boson couplings to 
fermions are severely reduced.  This is exactly what the combination of the CMS FP Higgs boson data indicates~\cite{CMSFPCOMB}. 
This implies, independently of the Higgs boson mass, that the Higgs production in gluon-gluon
fusion  $gg\to H$ (ggF), that is the dominant  SM-like Higgs production process at the LHC, is negligible for the FP Higgs boson. 
The dominant FP Higgs production mechanisms at the LHC are VBF and associate production with vector bosons, 
$VH,$ $V=W,\,Z,$ that are at least an order of magnitude smaller than the SM production cross section $\sigma(gg\to H).$
At the same time, due to the suppressed decay $H\to b\bar b$, the FP Higgs boson branching ratios to the gauge bosons,
$\gamma\gamma,$ $WW^*,$ $ZZ^*$ and $\gamma Z$ are strongly enhanced
for $m_H\lsim 140$ GeV. As a result, the production cross section times branching 
ratio, $\sigma\times BR,$ the quantity that is observable at the LHC, becomes strongly Higgs mass dependent for the FP Higgs boson. 
We note that in the case of  a heavier FP Higgs, the high mass region is not yet ruled out by the LHC~\cite{Gabrielli:2011bj}.

In this Letter we show that  for the FP Higgs boson with mass around 124~GeV the inclusive production cross section in VBF plus VH channels times
branching fraction into $\gamma\gamma$ is, within errors,  equal  to the SM Higgs boson $\sigma\times BR$ that is dominated by the ggF production. 
For the other channels, $WW^*,$ $ZZ^*,$ $Z\gamma,$ we predict a moderate  reduction of $\sigma\times BR$
for the FP Higgs.  Consequently,  the present inclusive LHC searches that reconstruct the Higgs invariant mass  
may accidentally support both the $m_H\approx 125$~GeV  SM-like Higgs and the FP Higgs bosons because
the fermionic channels $b\bar b$ and $\tau\bar\tau$ do not allow to discriminate between the two scenarios yet.
However,  the CMS $H\to \gamma\gamma$ searches~\cite{Cgg,CMSFP} do include 
also the exclusive VBF channel with  dijet tagging as one of the search categories, 
and the ATLAS inclusive search for FP Higgs~\cite{AFP} is optimized to diminish the ggF contribution.
Because the signal exceeds expectations in those searches, the present LHC results support the
possibility that the Higgs boson is fermiophobic.

The aim of this Letter is to emphasize that with a dedicated analysis optimized for the FP Higgs scenario the LHC
experiments will be able to test this scenario already this year,  possibly  providing a first measurement of the new physics scale $\Lambda$ 
connected to  the fermion masses generation.
In the VBF  production, the transverse momenta of the forward jets coming from  the scattered quarks balance 
the transverse momentum of the Higgs invariant mass system, that is  larger  than the typical Higgs momentum in the ggF
 process~\cite{Djouadi:2005gi}. This helps also for the background  suppression, as demonstrated in~\cite{AFP}.
These factors allow the experiments to discriminate between the two dominant production mechanisms.
 The first LHC searches~\cite{AFP,CMSFP}  prove that  the FP Higgs scenario can conclusively be tested with this year statistics
 surpassing the existing LEP~\cite{Heister:2002ub,Abreu:2001ib,Achard:2002jh,Abbiendi:2002yc},  
 Tevatron~\cite{TevatronNewHiggsWorking:2011aa} and the previous LHC bounds~\cite{Aold,Cold}.
 Indeed,  the LHC is much more sensitive to the FP Higgs scenario than to any other new physics scenario in which Higgs
 production is dominated by $gg\to H,$ for example supersymmetry.  To test  supersymmetric Higgs production one could need  
 to measure  a few percent level deviations~\cite{Djouadi:2005gj} from the SM prediction which may require years of running.

 The importance of testing the FP Higgs scenario at the LHC  goes far beyond 
 ruling out or ruling in one particular new physics scenario. Clearly, if the FP Higgs is ruled out, this implies that 
 $gg\to H$ is the main Higgs production mechanism as in the SM, and the Higgs Yukawa coupling to top quarks is indirectly measured.
 On the other hand, if the LHC experiments show that the presently preferred $\approx 125$~GeV Higgs boson is fermiophobic,
 our current understanding of electroweak symmetry breaking and flavor physics must be revised. All models with fundamental Yukawa couplings, 
 including the SM and the supersymmetric models, must be replaced with new mechanisms, \eg, technicolor.    
 One of the problems of the SM is a lack of dark matter that in the FP Higgs framework could be extra scalars that are stable 
 due to matter parity~\cite{Kadastik:2009dj}. 
  If the Higgs VBF and VH production processes dominate, 
 invisible FP Higgs boson decays~\cite{Eboli:2000ze,Godbole:2003it,Duhrssen:2004cv}  into dark matter scalars are enhanced at the LHC
 due to the increased branching fractions.
 This scenario may already have been hinted by the LHC $WW^*$ data~\cite{Mambrini:2011ik,Raidal:2011xk}.
\begin{figure}[htbp]
\begin{center}
\vskip 0cm
\includegraphics[width=0.43\textwidth]{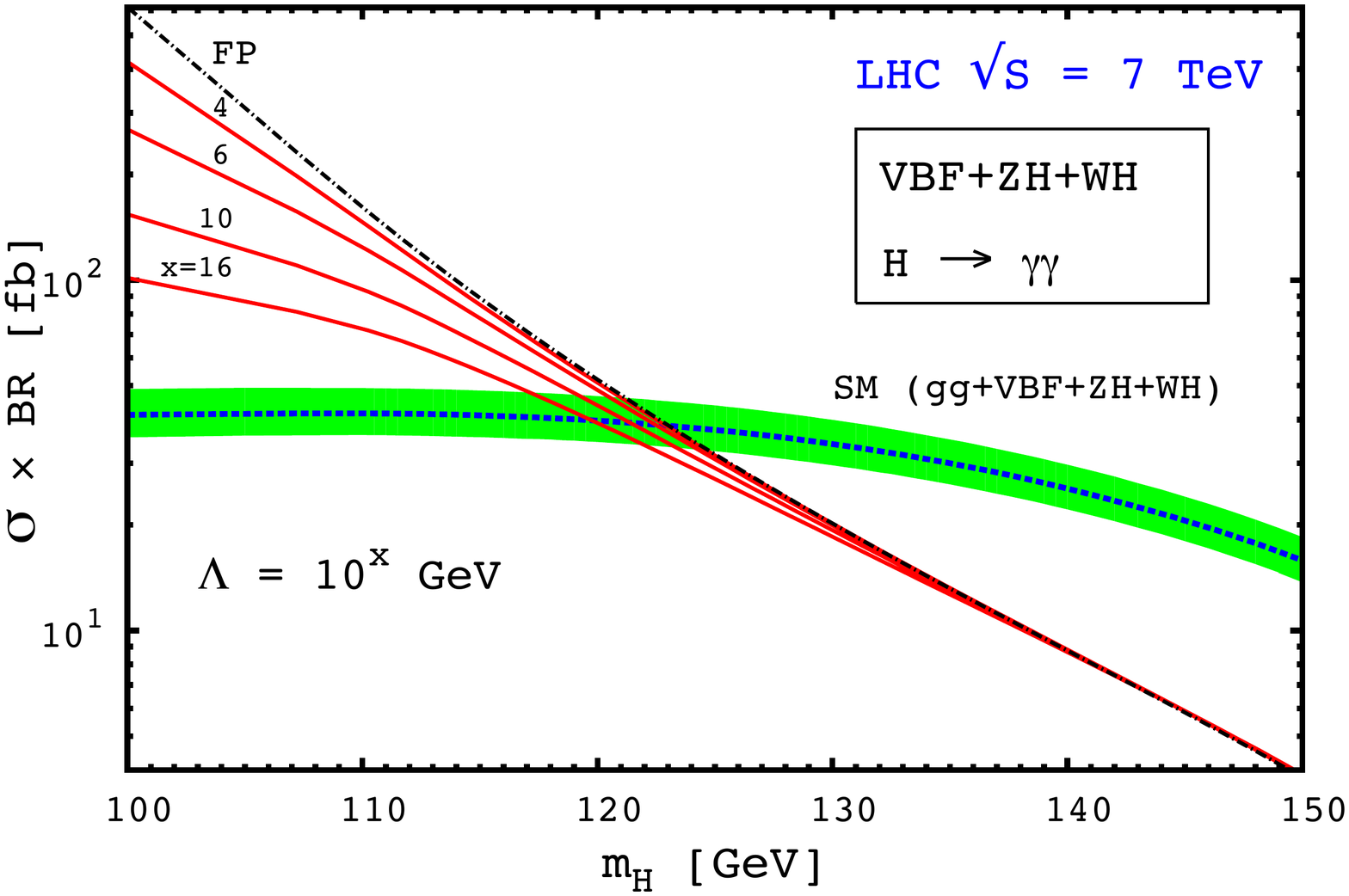}
\vskip -1cm
\includegraphics[width=0.43\textwidth]{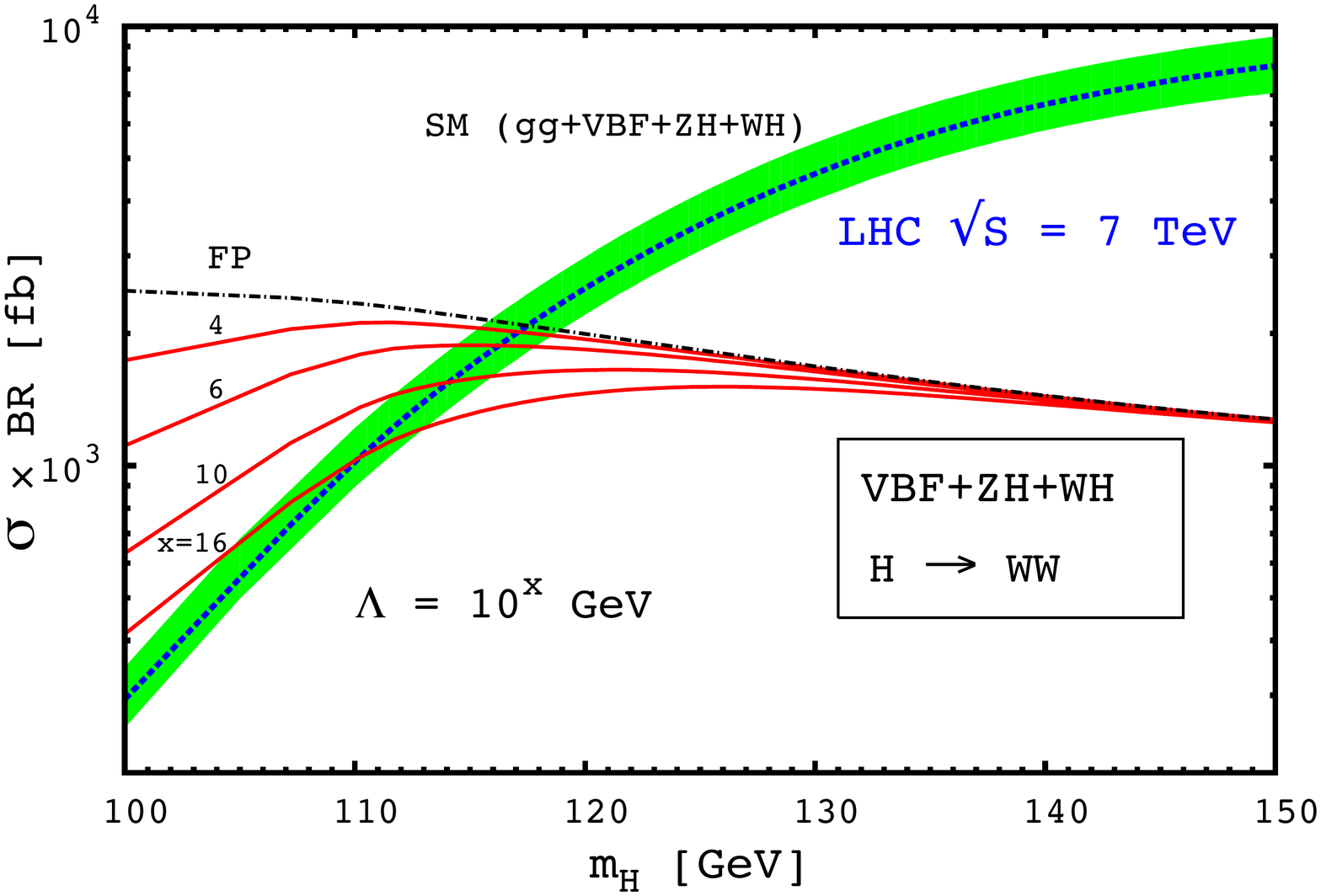}
\vskip -1cm
\includegraphics[width=0.43\textwidth]{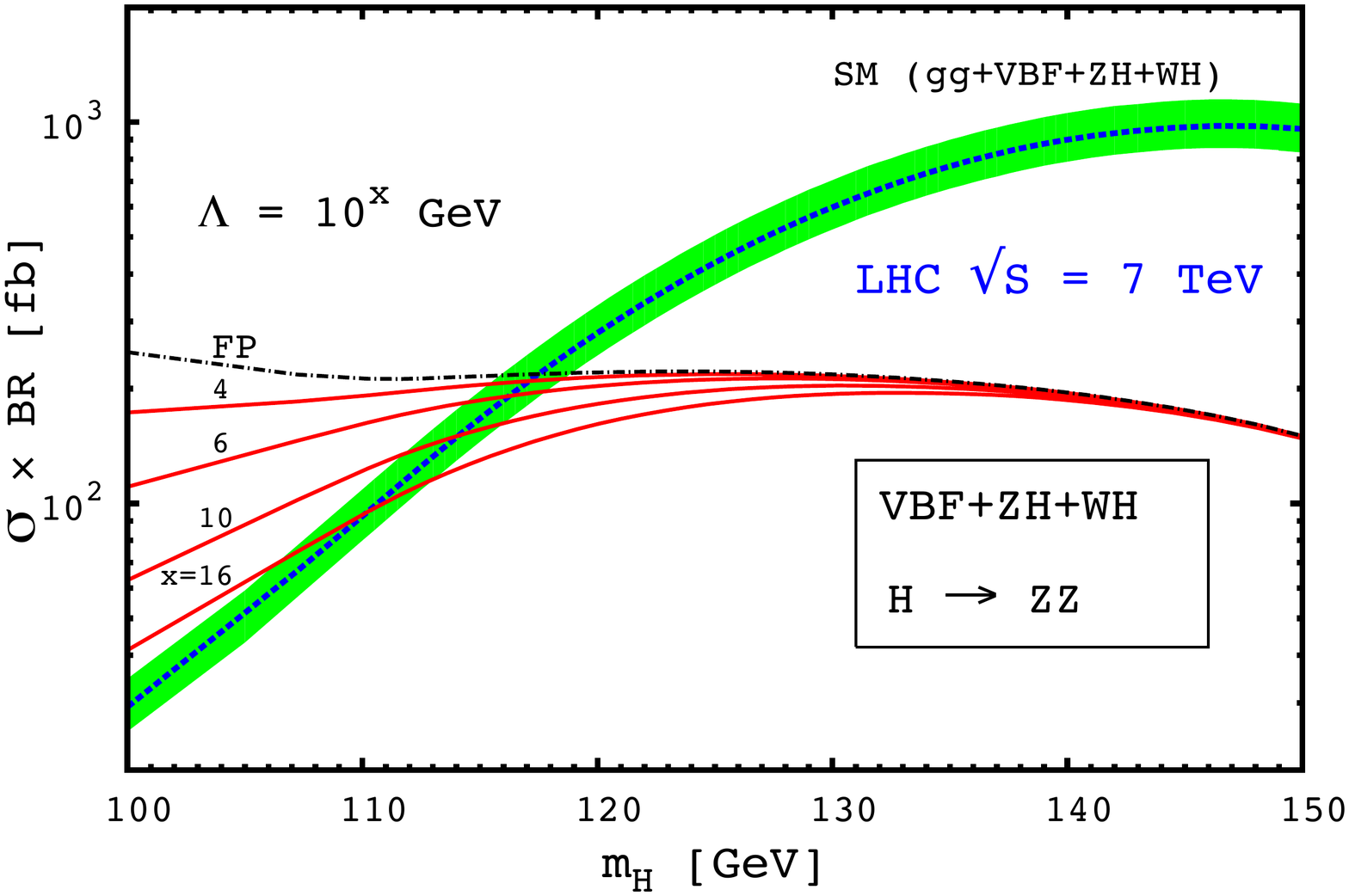}
\vskip -1cm
\includegraphics[width=0.43\textwidth]{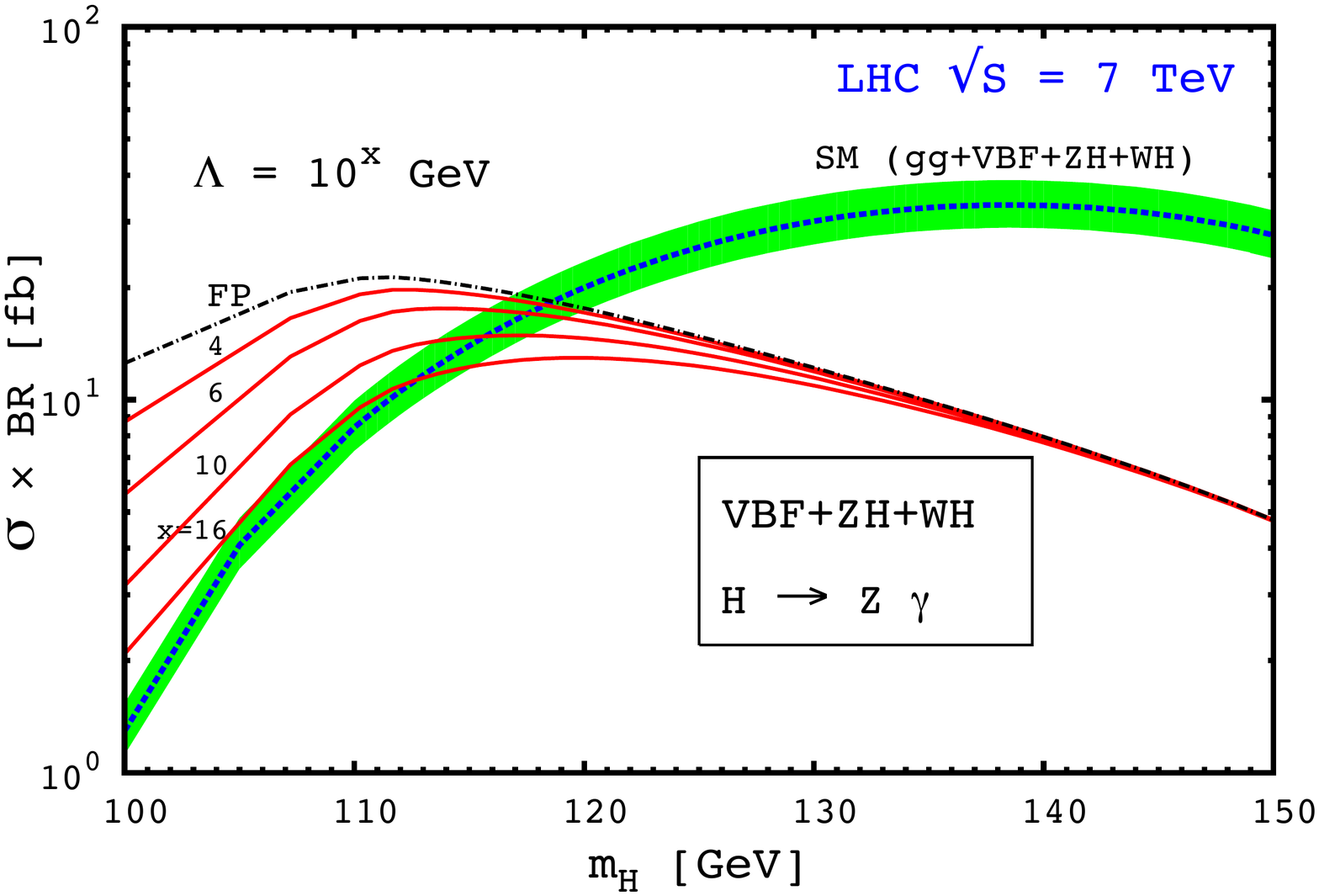}
\vskip -1cm
\caption{Dependence of the inclusive 
cross-sections times branching fraction for the FP Higgs boson decays into $\gamma\gamma,$ $WW^*,$ $ZZ^*$ and $Z\gamma$  
(from up to down)  on the Higgs mass $m_H$ at 7~TeV LHC.  Dash-dotted lines stand for the plain FP model, while red continuous lines 
represent the FP model with the inclusion of radiative corrections for several values of the new-physics scale $\Lambda=10^{4,6,10,16}$ GeV. 
The dotted lines present the central values of the SM Higgs inclusive production, together with the theoretical error shown by 
the green band.} 
\label{fig1}
\end{center}
\end{figure}

\mysection{A  fermiophobic Higgs boson at the LHC}
Here we present the inclusive FP Higgs boson production in the VBF, $ZH$ and $WH$ processes followed by  decays into gauge bosons.
To calculate the production cross sections and branching fractions in the FP Higgs scenario we include the radiative corrections due to the SM fermion masses 
into our analyses following~\cite{Gabrielli:2010cw}.  The radiative corrections depend logarithmically on the unknown new-physics scale $\Lambda$. 
We, therefore, treat this arbitrariness as a theoretical uncertainty on our predictions for $\sigma\times BR$ in the  FP Higgs scenario.

As we discussed above, our aim is to compare the FP Higgs signal with the SM-like Higgs signal at the LHC. 
We, therefore, use the state-or-art estimation~\cite{cx} of the inclusive SM Higgs production cross section in
ggF, VFB and $VH$ associate production channels to
compare our results with the SM predictions. In Fig.~\ref{fig1} we present our results for $\sigma \times BR$ 
in the $\gamma\gamma,$ $WW^*,$ $ZZ^*$ and $Z\gamma$ channels both for the FP Higgs boson and for the SM Higgs boson.
Dash-dotted lines stand for the pure FP model. The red continuous lines 
indicate the FP model with the inclusion of radiative corrections for several values of the new-physics scale $\Lambda=10^{4,6,10,16}$ GeV. 
The dotted line presents the central value of the SM Higgs inclusive production 
together with the theoretical error~\cite{cx} presented by  the green band.
As seen in the upper panel of Fig.~\ref{fig1}, the predictions for the FP Higgs model and for the SM Higgs in the $\gamma\gamma$ channel
coincide if the Higgs boson mass is around 123~GeV.  This value is close to the central value of the combined CMS Higgs signal.  
The ATLAS combined analyses prefers a somewhat higher Higgs boson mass, $m_H=126$~GeV.
 Since the $\gamma\gamma$ channel dominates the 
excess in both experiments, we conclude, based on the discussion above, that the FP Higgs boson mimics the SM Higgs boson in the present
searches. The two models can be discriminated by performing a dedicated search for the FP Higgs boson. 
The CMS exclusive VBF analyses in the $\gamma\gamma$ channel demonstrates that  this goal is achievable.

Going beyond the $\gamma\gamma$ signal,  we present  in the lower panels  of  Fig.~\ref{fig1} our predictions for the FP Higgs $\sigma \times BR$  
in the $WW^*,$ $ZZ^*$ and $Z\gamma$ channels. Those are predicted to be systematically lower than in the SM by a few tens of percent. 
This is one of our key predictions for the FP Higgs boson signal at the LHC that seems to be in qualitative agreement with the 
recently published  combined Higgs boson results. 
Thus the present experimental data may support, although inconclusively,  the FP Higgs production pattern over the SM.

\begin{figure}[t]
\begin{center}
\includegraphics[width=0.5\textwidth]{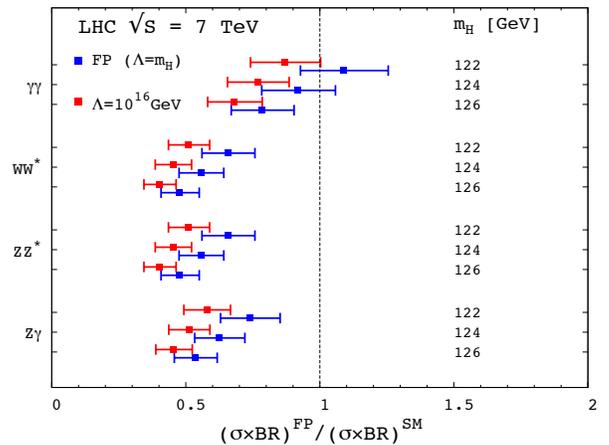}
\vskip -1cm
\caption{
Relative magnitudes of the FP Higgs prediction over a SM-like Higgs in different channels at the 7~TeV LHC for $m_H=122,\,124,\,126$~GeV.
The error bars correspond to the SM cross section uncertainties. The red (upper) and blue (lower) predictions show the theoretical 
errors associated with the new-physics scale $\Lambda.$ For LHC at  8~TeV,  the results are practically identical.
} 
\label{fig2}
\end{center}
\end{figure}

Based on the ATLAS and CMS results, we focus on the Higgs boson mass region between 122~GeV and 126~GeV. 
We  plot  in Fig.~\ref{fig2}  our predictions for $(\sigma \times BR)^{FP}/(\sigma\times  BR)^{SM}$ for the
different Higgs boson signatures assuming  $m_H=122,\,124,\,126$~GeV for two extreme values of the scale $\Lambda=m_H, 10^{16}$~GeV. 
The squares represent the central values of  predictions, while the error-bars take into account the uncertainty in the SM cross sections. The theoretical uncertainties in the FP Higgs predictions due to the unknown new-physics scale are shown by the effect of the change in  $\Lambda.$
We can see that the $\gamma\gamma$ rates can be completely compatible with the SM ones.
On the other hand, the other gauge boson channels show a depletion by a few tens of percents 
depending on the Higgs boson mass. 
 We have checked that the ratios in Fig.~\ref{fig2}
 are practically identical  for the 8~TeV LHC, since the dominant cross sections scale similarly with c.m. energy.
 
 Although it is too early to draw statistically relevant conclusions, one can be tempted  to compare our results in Fig.~\ref{fig2}
 with the CMS FP Higgs boson combination~\cite{CMSFPCOMB}. The latter reports a suppressed $H\to WW^*$ channel in VBF.
% in agreement with our predictions in Fig.~\ref{fig2}.  
 In addition, the global fit indicates a somewhat suppressed FP Higgs production 
 compared to the plain FP model predictions. The pattern observed by CMS could indeed be connected to a  large $\Lambda$ value.

\mysection{Impact of the results on  new physics scenarios}
Dedicated searches for the FP Higgs boson at the LHC imply non-trivial results on fundamental physics
independently of the outcome. If the FP Higgs boson will be ruled out by the LHC, 
the dominant Higgs production is determined to be ggF. This implies that the Higgs boson must have
SM-like Yukawa couplings to the top quark that can indirectly be measured. Consequently,  new physics scenarios
like supersymmetry and multi-Higgs models will be favored.

If the Higgs boson turns out to be fermiophobic, it breaks the electroweak symmetry but does not give  mass to quarks and leptons.
In this case our standard views on electroweak symmetry breaking and on flavor physics must be revised.
A particularly interesting question is what gives mass to the top quark. This would motivate studies of the top quark
couplings  at the LHC with the aim of finding unknown new physics that would be involved in this sector.

If the Higgs boson is fermiophobic, studying Higgs invisible decays becomes easier at the LHC, due to the enhanced branching ratios, and  may  probe Higgs 
couplings to dark matter. Therefore direct discovery of dark matter particles at the LHC will require less statistics than in the case of the 
SM-like Higgs boson.

\mysection{Conclusions}
We have shown that the recently published  inclusive ATLAS and CMS Higgs boson searches based on 2011 data do not 
discriminate between a SM-like Higgs boson and a FP Higgs boson. This is because for the Higgs boson mass 
presently favored by the LHC experiments, the Higgs production cross sections times branching fractions are 
accidentally similar in both scenarios. This happens despite of an order-of-magnitude difference in production cross sections and branching
fractions in the two scenarios. At the 7~TeV LHC we predict similar Higgs signals in the $\gamma\gamma$ channel  in both scenarios while
the $WW^*,$ $ZZ^*$ and $Z\gamma$ channels are predicted to be somewhat suppressed in the FP Higgs scenario.
Although not yet conclusively, the new ATLAS and CMS results support a FP Higgs boson with a new-physics scale $\Lambda$ possibly close to the GUT scale.
We urge the LHC experiments to perform a dedicated inclusive search for the FP Higgs boson in all channels that
should be able to discriminate between the two scenarios with data to be collected this year.
Ruling out or confirming the FP Higgs boson scenario will lay a path for  future new physics searches at the LHC,  
 giving information on the Higgs boson couplings   to fermions, gauge bosons and possibly also to dark matter.

{\it Acknowledgement.} We thank Andrey Korytov, Fabio Maltoni and Bill Murray for useful discussions. 
This work was supported by the ESF grants  8090,  MTT59, MTT60,   
by the recurrent financing SF0690030s09 project
and by  the European Union through the European Regional Development Fund.

\end{document}